# Bright Luminescent Surface States on the Edges of Wide-bandgap Two-dimensional Lead Halide Perovskite

Yanan Wang,[1,3,†,¥] Chong Wang,[2,3,†] Xinghua Su,[4,3,†] Viktor G. Hadjiev,[5,6] Hector A Calderon Benavides[7] Yizhou Ni,[5,8] Md Kamrul Alam,[9] Francisco C. Robles-Hernandez,[10,3] Yan Yao,[3,5,9] Shuo Chen,[5,8] Zhiming Wang,[1,*] Jiming Bao[3,1,9,*]

[1]Institute of Fundamental and Frontier Sciences, University of Electronic Science and Technology of China, Chengdu, Sichuan 610054, China

[2]Institute of Optoelectronic Information Material, School of Material Science and Engineering, Yunnan University, Kunming, Yunnan 650091, China

[3]Department of Electrical and Computer Engineering, University of Houston, Houston, Texas 77204, United States

[4]School of Materials Science and Engineering, Chang'an University, Xi'an, Shaanxi 710061, China

[5]Texas Center for Superconductivity, University of Houston, Houston, TX 77204, United States

[6]Department of Mechanical Engineering, University of Houston, Houston, TX 77204, United States

[7]Instituto Politecnico Nacional, ESFM-IPN, UPALM, Departamento de Física, Mexico CDMX 07338, Mexico

[8]Department of Physics, University of Houston, Houston, Texas 77204, United States

[9]Materials Science & Engineering, University of Houston, Houston, Texas 77204, United States

[10]Mechanical Engineering Technology, University of Houston, Houston, TX 77204, United States

[†]Equal contributions

*Corresponding author. E-mail: jbao@uh.edu (J.M.B.); zhmwang@uestc.edu.cn (Z.M.W.)

[¥]Current address: Electrical Engineering & Computer Science, Case Western Reserve University, Cleveland, Ohio 44106, United States

## Abstract

Three-dimensional lead halide perovskites have surprised people for their defect-tolerant electronic and optical properties, two-dimensional lead halide layered structures exhibit even more puzzling phenomena: luminescent edge states in Ruddlesden-Popper perovskites and conflicting reports of highly luminescent versus non-emissive $CsPb_2Br_5$. In this work, we report the observation of bright luminescent surface states on the edges of $CsPb_2Br_5$ microplatelets. We prove that green surface emission makes wide-bandgap single crystal $CsPb_2Br_5$ highly luminescent. Using polarized Raman spectroscopy and atomic-resolution transmission electron microscopy, we further prove that polycrystalline $CsPb_2Br_5$ is responsible for the bright luminescence. We propose that these bright edge states originate from corner-sharing clusters of $PbBr_6$ in the distorted regions between $CsPb_2Br_5$ nanocrystals. Because metal halide octahedrons are building block of perovskites, our discoveries settle a long-standing controversy over the basic property of $CsPb_2Br_5$ and open new opportunities to understand, design and engineer perovskite solar cells and other optoelectronic devices.

Lead halide perovskites have provided us not only an ideal material platform to realize the dream of high-efficiency solar cells and many other optoelectronic devices, but also a wide range of structures to explore unusual fundamental sciences[1, 2, 3, 4]. Depending on spatial configurations, lead halide octahedrons can form structures from three dimensional (3D) all the way to 0D perovskites[5]. While still not completely understood, the superior optoelectronic properties of perovskites are believed to originate from their immunity to defects and lack of non-radiative deep level traps. More surprising is recent observation of highly emissive deep-level states on the edges of 2D Ruddlesden-Popper (R-P) $(BA)_2(MA)_{n-1}Pb_nI_{3n+1}$ perovskites and subsequent utilization of such edge state to further enhance the efficiency of solar cells[6]. However, the nature of the edge states is not clear, and evens their chemical composition and microscopic structure have not been identified[6].

$CsPb_2Br_5$ is another type of layered lead halide structure with edge-shared $PbBr_6$ octahedrons forming flat sheets and Cs as spacing layers. 2D $CsPb_2Br_5$ has also attracted a lot of attention recently due to many conflicting reports on its luminescence although it was synthesized and studied long ago[7, 8]. Zhang et al. were the first to report the beneficial effect of $CsPb_2Br_5$ to a 3D all-inorganic perovskite $CsPbBr_3$: the attachment of $CsPb_2Br_5$ nanoparticles to $CsPbBr_3$ nanocrystals enhanced photoluminescence (PL) of $CsPbBr_3$ by several folds and external quantum efficiency of $CsPbBr_3$ light-emitting diodes (LEDs) by 50%[9]. Shortly after that, Wang and co-workers reported nearly 90% quantum efficiency of pure $CsPb_2Br_5$ nanoplatelets and subsequently expanded their emission wavelength to whole visible spectrum using ion exchange with I and Cl[10]. But these claims of highly luminescent $CsPb_2Br_5$ have been met with skeptics. Li *et al*. observed no PL at all and further proved it as an indirect wide bandgap semiconductor with theoretical support from density functional theory (DFT) simulation[11]. Since then, the controversy has remained, while some groups still reported strong visible photoluminescence, high efficiency LEDs, photodetectors, and even lasing action in $CsPb_2Br_5$ microplates[12, 13, 14, 15, 16, 17], others supported the opposite, believing PL due to embedded $CsPbBr_3$[18, 19]. Many groups were aware of this controversy but were not able to support either of these two opposing claims[20, 21, 22].

Edges are intrinsic surfaces of two-dimensional materials, and typically are sources for defects that can greatly degrade device performances. Here we report the observation of bright photoluminescence from the edges of $CsPb_2Br_5$ microplatelets and prove that polycrystalline $CsPb_2Br_5$ is responsible for luminescent edge states of non-luminescent wide-bandgap single crystal $CsPb_2Br_5$. This discovery is made possible by our unique synthesis and characterization techniques. Pure water is the only chemical needed for the synthesis of $CsPb_2Br_5$ from $CsPbBr_3$. $CsPb_2Br_5$ microplatelets with different edge morphologies are obtained such that edge emissions can be spatially located and distinguished from the body emission. Polarization-dependent micro-Raman in conjunction with micro-PL and atomic-resolution transmission electron microscopy (TEM) allows us to correlate luminescent sources to the underlying crystal structure. In-situ PL imaging of water-induced $CsPbBr_3$ to $CsPb_2Br_5$ phase transformation and combined Raman/PL of water-immersed $CsPb_2Br_5$ further exclude $CsPbBr_3$ as a luminescent source. Our discoveries settle the debate on the basic properties of $CsPb_2Br_5$ and pave the way for further exploration, understanding and device applications of a wide range of lead halide perovskites.

$CsPbBr_3$ powders were first synthesized using a modified method by mixing 0.5 M $Pb(CH_3COO)_2 \cdot 3H_2O$ and 1 M CsBr in 48% HBr solution at room temperature[23, 24]. $CsPb_2Br_5$ was then synthesized by simply dropping $CsPbBr_3$ micropowders in a large quantity of water in a flask (20-50 times more water in mass) at room temperature. Orange $CsPbBr_3$ quickly turned white and precipitated at the bottom of the flask. The white precipitates can be taken out and dried for further study. The very pure phases of initial $CsPbBr_3$ and precipitated $CsPb_2Br_5$ were verified by their clean XRD patterns.

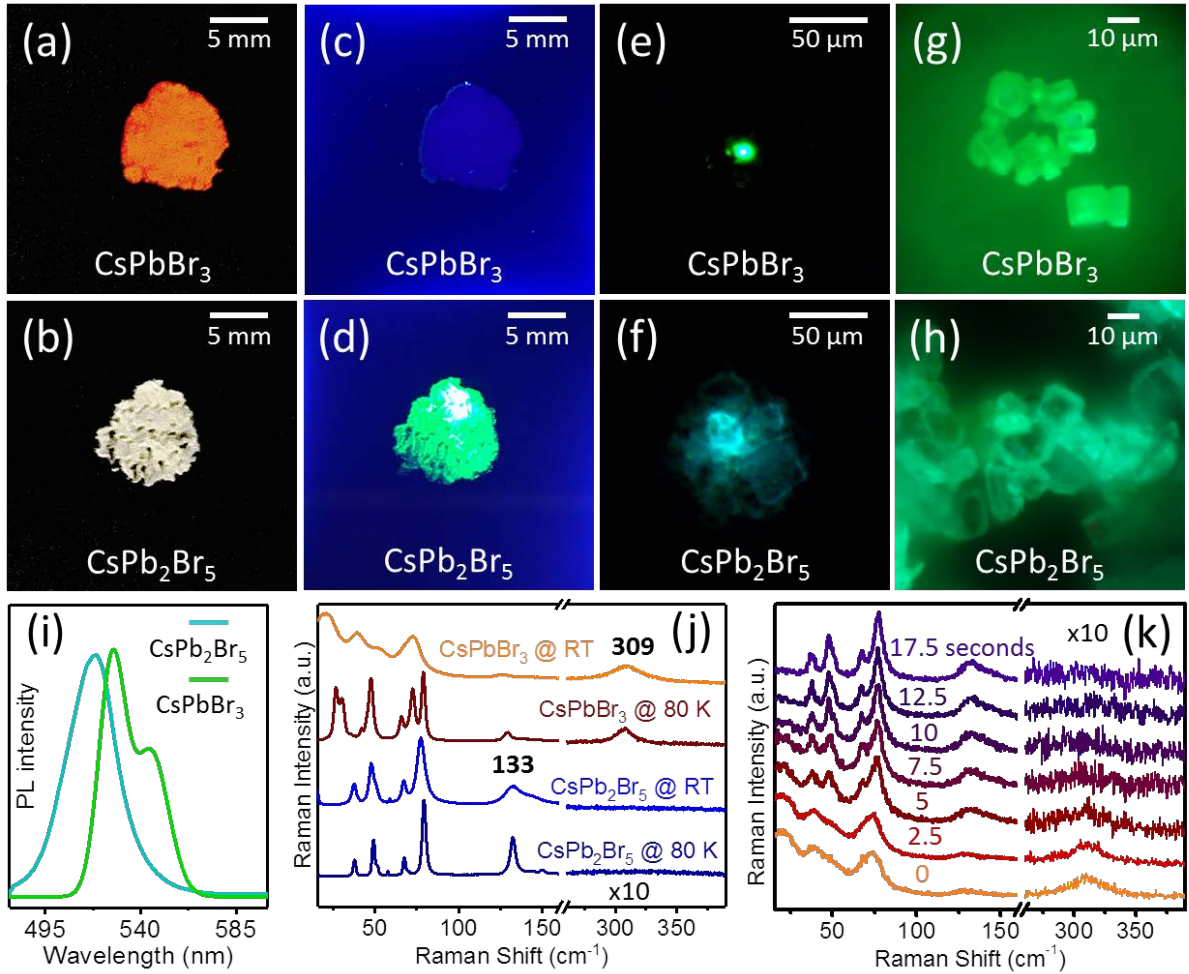

Fig. 1. Optical and luminescence images of $CsPbBr_3$ and $CsPb_2Br_5$ as well as their Raman and photoluminescence spectra. (a-d) Pictures of $CsPbBr_3$ and $CsPb_2Br_5$ under (a-b) ambient light and (c-d) 365-nm UV illuminations. (e-h) Microscopic photoluminescence images of $CsPbBr_3$ and $CsPb_2Br_5$ under (e-f) 473-nm laser (g-h) focused 365-nm UV illuminations. (i) Typical photoluminescence spectra of $CsPbBr_3$ and $CsPb_2Br_5$ powders. (j) Raman spectra of $CsPbBr_3$ and $CsPb_2Br_5$ powders at room temperature and 80 K. (k) Evolution of Raman spectrum when water was added to $CsPbBr_3$.

Figs. 1a-b show photographs of $CsPbBr_3$ and $CsPb_2Br_5$ powders. The color change from orange to white not only confirms a crystal phase transformation, but also indicates a potentially big increase in bandgap. However, when illuminated by a UV light (Figs. 1c-d), $CsPb_2Br_5$ powders

emit bright green light. In contrast, CsPbBr$_3$ remains almost dark, emitting weak light. The difference is still distinguishable under an optical microscope when excited by a 473-nm laser (Figs. 1e-f): a strong green emission is only observed in CsPbBr$_3$ from the laser excitation spot, while for CsPb$_2$Br$_5$, a large region surrounding the laser excitation spot becomes bright. More importantly, it appears that bluish green emission comes from the edges of individual CsPb$_2$Br$_5$ crystals. This difference in luminescence images can be seen more clearly when both CsPbBr$_3$ and CsPb$_2$Br$_5$ powders are under a stronger UV illumination (Figs. 1g-h): light comes from whole CsPbBr$_3$ crystals uniformly while only edges of CsPb$_2$Br$_5$ becomes bright.

The difference in emission wavelength is also revealed by their spectra in Fig. 1i. CsPbBr$_3$ exhibits two peaks at 528 and 542 nm[25], but CsPb$_2$Br$_5$ has a broad peak blue shifted from the PL of CsPbBr$_3$. This phase change was confirmed by Raman spectroscopy[26, 27, 28]. Fig. 1j shows Raman spectra of CsPbBr$_3$ and CsPb$_2$Br$_5$ at room temperature and 80 K. The spectra of the two compounds can be distinguished easily at room temperature by the characteristic sharp lines of CsPb$_2$Br$_5$ at low wavenumbers versus the two-phonon mode of CsPbBr$_3$ at 309 cm$^{-1}$ and by strong double peaks of CsPbBr$_3$ at 25 cm$^{-1}$ at low temperature[26, 27, 28]. The fast transition from CsPbBr$_3$ to CsPb$_2$Br$_5$ can also be seen from in-situ Raman in Fig. 1k.

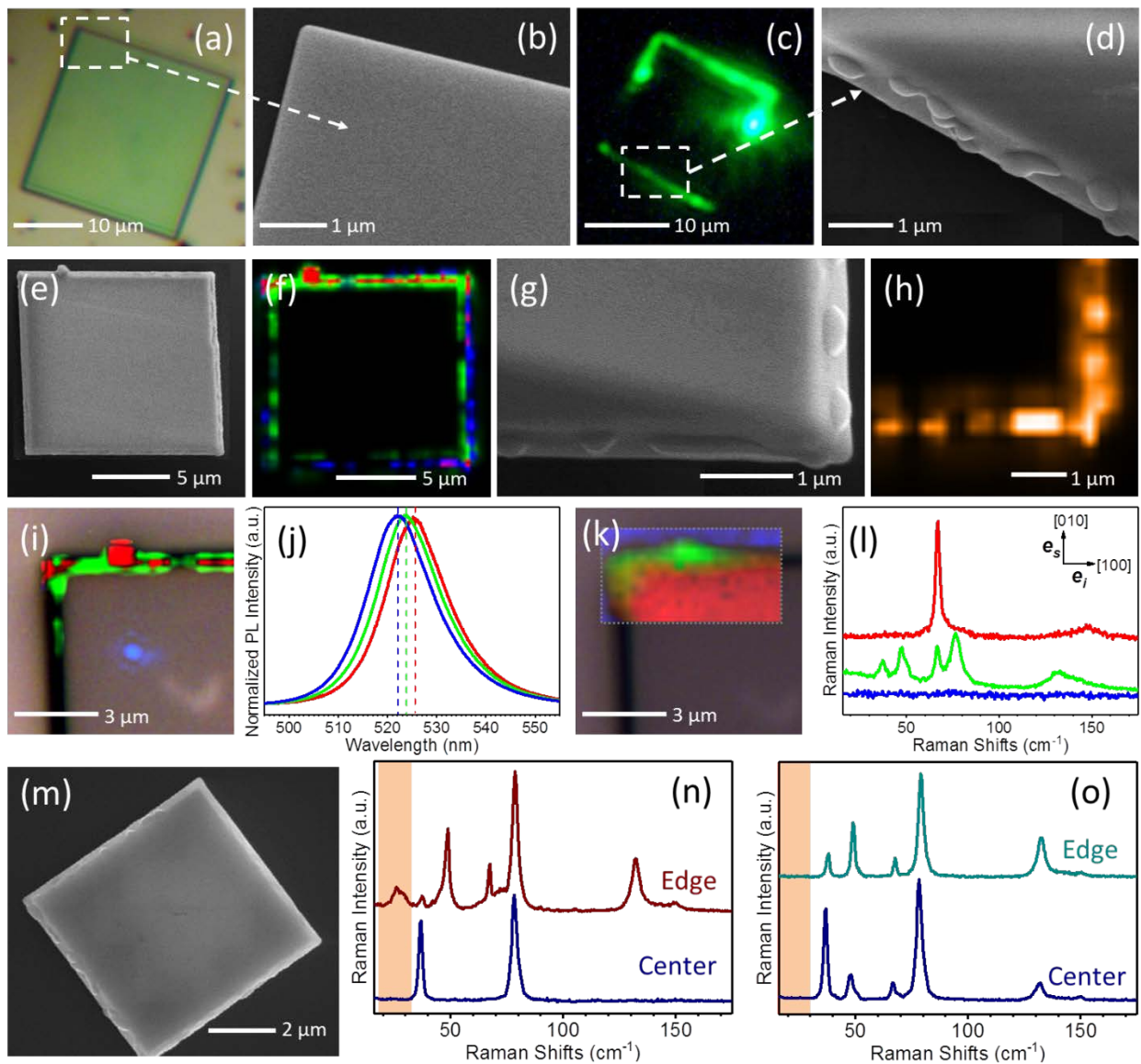

Fig. 2. Identification of edge emission in $CsPb_2Br_5$ microplatelets using PL, SEM and Raman. (a) Optical and (b) SEM images of a $CsPb_2Br_5$ platelet that shows no PL. (c) Photoluminescence image of a 2$^{nd}$ $CsPb_2Br_5$ platelet when excited by a 473-nm laser in the center. (d) SEM of bright edge of the platelet in (c). (e-l) SEM, PL and Raman characterizations of a 3$^{rd}$ $CsPb_2Br_5$ platelet. (e) SEM and (f) confocal photoluminescence mapping. (g) SEM image and (h) photoluminescence intensity mapping of low right corner of the platelet. (i-l) Confocal photoluminescence and Raman mappings of top left corner of the platelet. (j) Three PL modes used for PL imaging in (f) and (i). (l) Three Raman

modes for the mapping in (k). (m-n) SEM and low-temperature Raman indicates co-existence of $CsPbBr_3$ and polycrystalline $CsPb_2Br_5$ in a luminescent bump of a 4$^{th}$ $CsPb_2Br_5$ platelet. (o) The absence of $CsPbBr_3$ Raman in low-temperature Raman scattering of a luminescent bump in a 5$^{th}$ platelet.

The apparent display of bright edges doesn't necessarily guarantee that light emission originates from the edges, it could simply be a result of edge light scattering and wave guiding of photoluminescence in $CsPb_2Br_5$ platelets as can be found in a luminescent solar concentrator[29]. To find out the exact source for the edge emission, we prepared $CsPb_2Br_5$ platelets with very flat and clean top surfaces. Square-shaped thin microplatelets were scooped from the surface of $CsPb_2Br_5$ synthesis water solution and dried by $N_2$ blow in air[8]. Two types of platelets were observed under laser or UV excitation. The first type showed no photoluminescence at all. Figs. 2a-b show optical and SEM images of such "dark" platelet: both edges and top surface are extremely smooth. Fig. 2c shows the other type of platelet where some locations of edges are bright but the body of the platelet is dark under laser excitation. Fig. 2d reveals that bright edges are decorated with individual nanobumps while dark edges are smooth without any bumps as in Fig. 2b. These new observations lead us to conclude that $CsPb_2Br_5$ is a wide bandgap semiconductor and transparent to visible light; bright edge emissions come from nanobumps on the edges.

To identify the structure and composition of the emitting edges, we combined micro-PL and micro-Raman to probe individual bumps. Figs. 2e-j show SEM and confocal PL mapping of a $CsPb_2Br_5$ platelet, a good overlap between PL mapping and SEM image confirms these bumps as the source for the edge emission. To further identify the bumps and their relationship to the $CsPb_2Br_5$ platelets, we performed a Raman mapping of the same top-left corner of the platelet as the one shown in the PL mapping in Fig. 2i. Here a 632.8-nm laser was used to eliminate strong PL background. In addition, Raman spectra were recorded in a scattering configuration that selects only two $B_{2g}$ phonon lines in good single crystals[28]. The Raman mapping in Fig. 2k reveals two distinct regions: a green region from the edge defined by a complete (depolarized) Raman spectrum (green) in Fig. 2(l) and red inner region characterized by two $B_{2g}$ peaks. The strongly polarized $B_{2g}$-spectra in the inner region indicate the good single crystal quality of the $CsPb_2Br_5$ platelet. Conversely, the emergence of all active modes indicates the presence of multiple $CsPb_2Br_5$ domains with different crystal orientations, i.e., polycrystalline $CsPb_2Br_5$. An excellent spatial and intensity overlap between Raman and PL mappings implies that polycrystalline $CsPb_2Br_5$ is the main body of PL active bumps. Figs. 2m-o shows low temperature Raman spectra from similar bumps in two other platelets. A weak $CsPbBr_3$ Raman feature at 25 cm$^{-1}$ can be seen in Fig. 2o, indicating co-existence of $CsPbBr_3$ and polycrystalline $CsPb_2Br_5$, while no $CsPbBr_3$ Raman is observed in Fig. 2o.

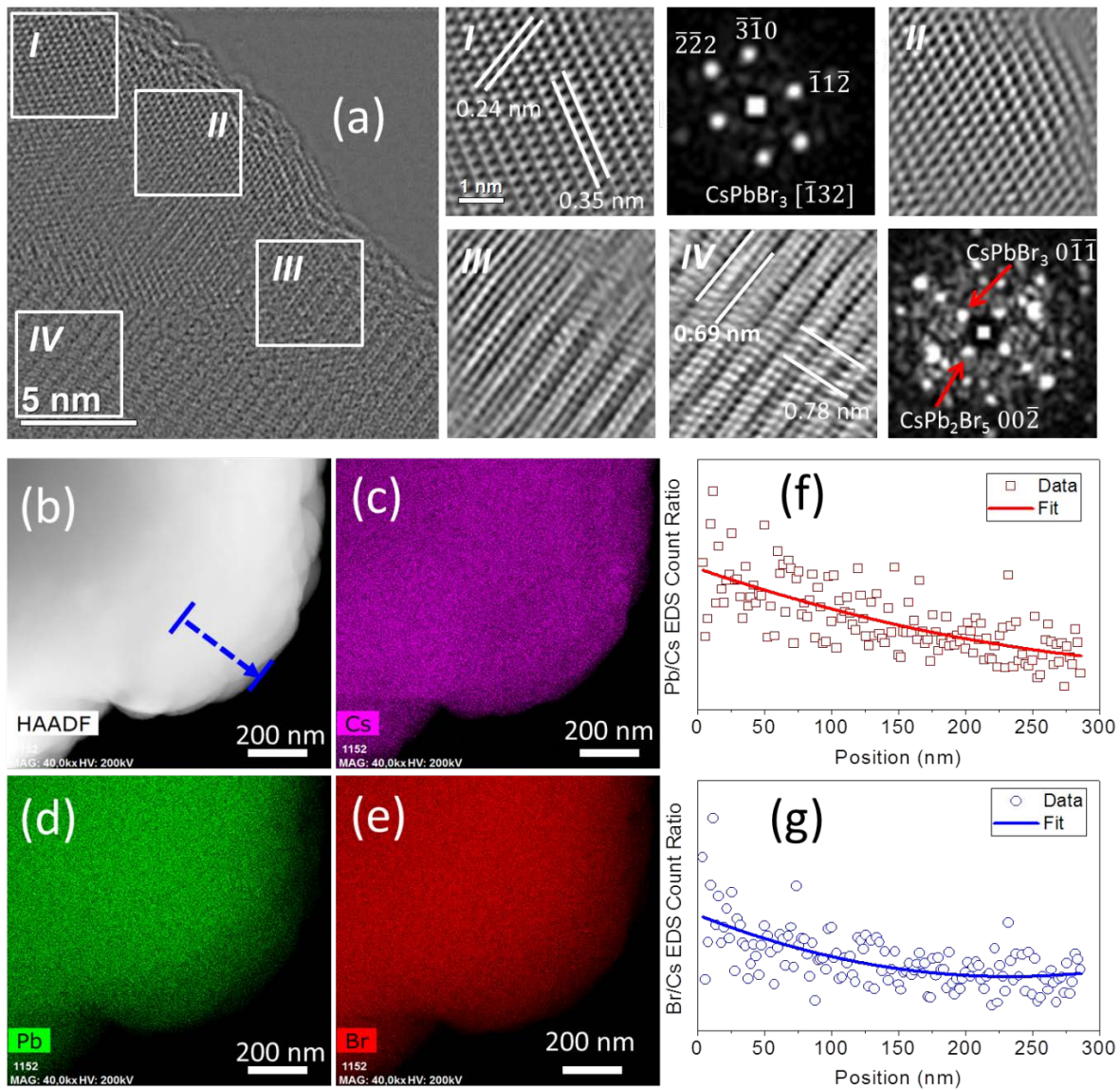

Fig. 3. Transmission electron microscopy (TEM) and energy-dispersive X-ray spectroscopy (EDS) investigations of a bump in a $CsPb_2Br_5$ platelet. (a) TEM phase image of the tip of the bump. Close-up and filtered sections of the phase images as indicated by regions I-IV; FFT electron diffraction patterns of the lattice in I and IV. (b) High angle annular dark field (HAADF) scanning TEM image of the bump. (c-e) EDS Cs, Pb and Br elemental mappings of the bump. (f-g) Pb/Cs and Br/Cs EDS count ratios along the line indicated in (b) and based on the Cs, Pb and Br mappings from (c) to (e).

The existence of both $CsPbBr_3$ and $CsPb_2Br_5$ in the same bumps can be most directly observed with atomic-resolution low dose aberration-corrected TEM and exit wave reconstruction (EWR) [30, 31]. The use of low dose allows preserving the synthesized sample without beam alteration and thus the corresponding images represent the genuine structure. A dose of 30 $e/Å^2 s$ or lower was used here. Fig. 3a shows phase image of a bump near the surface. Also shown are selected regions I through IV and electron diffraction patterns of regions I and IV. Clearly, this whole region is not a single crystal lattice. The regions I and II can be clearly identified as $CsPbBr_3$, but regions III and IV exhibit distorted lattice, can be hardly identified as pure $CsPbBr_3$ or $CsPb_2Br_5$. For example, region IV can only be identified as a mixture of $CsPbBr_3$ and $CsPb_2Br_5$. Because the base of the bump is single crystal $CsPb_2Br_5$, we believe that TEM reveals a gradual phase transition from $CsPb_2Br_5$ to $CsPbBr_3$. Such transition can also be qualitatively confirmed by energy-dispersive X-ray spectroscopy (EDS) elemental mapping in Fig. 3b-e. From the inner $CsPb_2Br_5$ region to the surface, Cs intensity remains almost constant, but Pb shows a rapid drop near the surface. Br intensity also drops, but not as fast as Pb near the surface. Such relative ratios of Pb and Br to Cs EDS counts are confirmed by analysis shown in Figs. 3f-g. These observations agree with the expectation from the transition from $CsPb_2Br_5$ to $CsPbBr_3$: Pb/Cs ratio will decrease from 2 in $CsPb_2Br_5$ to 1 in $CsPbBr_3$, while the Br/Cs ratio should decrease from 5 in $CsPb_2Br_5$ to 3 in $CsPbBr_3$.

The identification of small amounts of $CsPbBr_3$ by TEM agrees with the Raman study in Fig. 2n, where Raman signatures of both $CsPbBr_3$ and $CsPb_2Br_5$ are observed. The absence of $CsPbBr_3$

Raman in Fig. 2o certainly indicates a negligible amount of $CsPbBr_3$ compared to $CsPb_2Br_5$. To obtain a lower limit of $CsPbBr_3$, we chose lasers with energy close to the PL peak so that Raman can be resonantly enhanced. An additional benefit of this approach is that photoluminescence and Raman can be captured in the same spectrum so that the contribution of $CsPbBr_3$ to the photoluminescence can be more accurately quantified. Fig. 4a shows a spectrum of the big bump in Figs. 2e, 2f and 2i by a 532-nm laser: a $CsPbBr_3$ Raman line at 309 $cm^{-1}$ can be clearly seen on the strong PL shoulder, the Raman of $CsPb_2Br_5$ at 133 $cm^{-1}$ appears relative weak due to even stronger PL background. This combined PL and Raman spectrum reveals a possible contribution of $CsPbBr_3$ to the PL and proves itself as a highly sensitive technique compared to a quick Raman mapping in Figs. 2k-l. Similar combined PL and Raman can also be achieved by a shorter wavelength laser. Fig. 4b shows a spectrum from a bump of another $CsPb_2Br_5$ microplatelet excited by a 473-nm laser. Despite a very strong luminescence, a clear $CsPbBr_3$ Raman peak at 309 $cm^{-1}$ can still be seen in the close-up view.

With this sensitive combined Raman/PL technique, we now show more cases that $CsPbBr_3$ alone cannot explain the bright photoluminescence, and polycrystalline $CsPb_2Br_5$ is carrying another highly luminescence agent even though its single crystal is totally transparent. Let's first take another look at $CsPbBr_3$ to $CsPb_2Br_5$ transition in water. Figs. 4c-f show in-situ PL imaging of the transition under the UV illumination. A big $CsPbBr_3$ single crystal can be easily recognized by its green body emission. The crystal was dissolved into several smaller pieces and gradually disappeared after several minutes. In the meantime, bright spots with bluish green color began to emerge from previous dark regions. This observation has given us very important information: these bright spots are not the initial $CsPbBr_3$ and cannot be made of $CsPbBr_3$, they must be polycrystalline $CsPb_2Br_5$. This assumption can be quickly confirmed. Figs. 4g-j show photoluminescence and Raman from these bright spots in water. A strong Raman peak at 133 $cm^{-1}$ confirms $CsPb_2Br_5$, no Raman peak of $CsPbBr_3$ at 309 $cm^{-1}$ is found. Note that the Raman spectrum in Figs. 4h-i is similar to that in Fig. 4b in terms of dynamic range and signal to noise ratio, if $CsPbBr_3$ is responsible for the bright emission, it must exhibit the characteristic Raman peak. The same conclusion can be made from the comparison between Fig. 4j and 4a when PL and Raman were excited by the 532-nm laser.

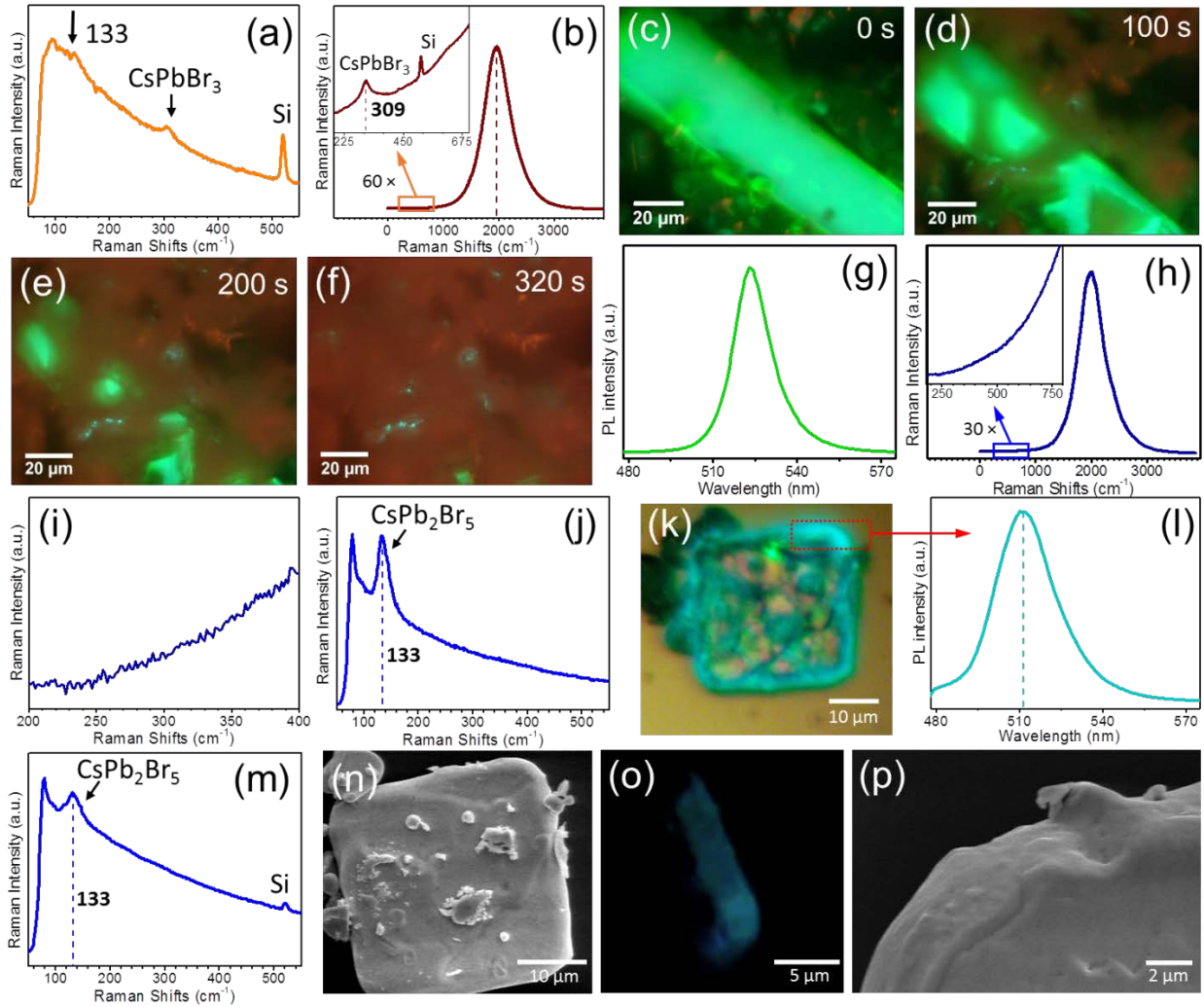

Fig. 4. PL and Raman spectroscopy of polycrystalline $CsPb_2Br_5$. (a) Combined Raman and PL spectrum of the big bump of the $CsPb_2Br_5$ platelet in Figs. 2e, 2i and 2k excited by a 532-nm laser. (b) Combined Raman and PL spectrum of a bump of a 5th $CsPb_2Br_5$ platelet excited by a 473-nm laser. (c-f) In-situ PL imaging of the dissolution of a large $CsPbBr_3$ crystal and formation of bright luminescent polycrystalline $CsPb_2Br_5$ spots. (g) Typical PL spectrum of a bright spot of $CsPb_2Br_5$ in water. (h-j) Combined Raman and PL spectra of a luminescent spot of $CsPb_2Br_5$ in water excited by (h-i) the 473-nm laser and (j) the 532-nm laser. (i) and inset of (h): zoomed-in spectrum near 309 cm$^{-1}$. (k) PL imaging, (l) PL spectrum, (m) Raman spectrum and (n) SEM of a $CsPb_2Br_5$ platelet. PL was excited by 473-nm laser and Raman was excited by 532-nm laser. (o) PL and (p) SEM image of edge of another $CsPb_2Br_5$ platelet.

We applied the same combined Raman/PL technique to the bright $CsPb_2Br_5$ in Figs. 1f and 1h again. Fig. 4k shows the PL image of a $CsPb_2Br_5$ crystal. As expected, this edge emission has a broad shorter wavelength centered at 511 nm (Fig. 4l). Raman spectrum from the same spot exhibits a strong $CsPb_2Br_5$ peak at 133 $cm^{-1}$ but without any sign of $CsPbBr_3$ peak at 309 $cm^{-1}$ (Fig. 4m). Based on the same reason, we conclude that the contribution of $CsPbBr_3$ to the photoluminescence is negligible, in other words, polycrystalline $CsPb_2Br_5$ is responsible for the strong PL. This shorter wavelength emission and lack of $CsPbBr_3$ is due to different edge morphology: SEM in Fig. 4n shows relatively smooth edges without large bumps. A direct edge emission and edge morphology can be found in Figs. 4o-p.

Now we have three types of $CsPb_2Br_5$ that show edge emissions: bumps on the edges of platelets (Fig. 2), immersed in water (Fig. 4c-j), and no bump but with relatively "rough" surfaces (Fig. 1f-h, Fig. 4k-p). Their different morphologies and compositions are due to slightly different drying processing. For the platelets with bumps in Fig. 2, they were scooped with a silicon substrate from the water surface and were blow dried by $N_2$ after they emerged to the surface from the precipitates at the bottom. In this case, some water will be left on the edges, and metal halide ions in the water will precipitate as water dries up. $CsPbBr_3$ nanocrystals might form on the top of polycrystalline $CsPb_2Br_5$; because $CsPbBr_3$ has a high solubility in water, it cannot crystalize until the platelet is dried. For this reason, the bright luminescent spots must come from polycrystalline $CsPb_2Br_5$ in water, in agreement with the combined Raman/PL investigation. On the other hand, if $CsPbBr_3$ nanocrystals can be embedded in $CsPb_2Br_5$, we should already have observed them inside $CsPb_2Br_5$ platelets and found their corresponding body emission[32, 33]. $CsPb_2Br_5$ platelets with relatively rough surface but without bumps are white precipitates, they were collected and then washed by diethyl ether for three times before they were dried in a furnace at 60 degrees for 12 hours. Rinsing with diethyl ether is a typical procedure to remove residual water from a crystal surface; as a result, no water droplets were left on the edges and no bumps were formed.

Let us discuss the possible causes for PL in polycrystalline $CsPb_2Br_5$. Many different types of defects can be created on the edges between the boundaries of $CsPb_2Br_5$ nanocrystals, including vacancies, interstitials and distorted bonds. Br vacancies can be excluded because a recent DFT study shows that Br vacancies in $CsPb_2Br_5$ monolayers create shallow levels at 0.12 – 0.25 eV below the conduction band[34]. Self-trapped excitons are another contender but they can also be excluded because the laser energies are too small to create free excitons in $CsPb_2Br_5$. We believe that clusters of corner sharing $PbBr_6$ octahedrons on the edges are responsible for the bright luminescent edge states. The bandgaps of perovskites are mainly determined by lead halide octahedrons and their framework. The distorted lattices or variation of the octahedral tilting angles between octahedrons in these macro-molecules made of individual $PbBr_6$ molecules will result in a slightly larger bandgap than that of fully connected $PbBr_6$ network, i.e., $CsPbBr_3$[35].

Our discovery and understanding of luminescent states on the edges of wide bandgap $CsPb_2Br_5$ single crystals explain many seemingly conflicting observations. For large and even millimeter size crystals, transparent non-emissive plates of $CsPb_2Br_5$ are more convenient to observe than possible edge emission; for small crystals especially submicron-sized particles or platelets, it is difficult to distinguish edge emission from body emission. In some cases, $CsPb_2Br_5$ could have potential "contamination" from $CsPbBr_3$ nanocrystals[13, 14, 15]. More research is clearly needed to verify the origin of luminescent deep level edge states. Metal halide octahedrons are the basic building blocks of perovskites with any dimensionality and understanding properties of clusters of corner sharing octahedrons is essential to the understanding of defects and defect states of any perovskites. Because such luminescent properties have already found applications in making high efficiency defect-tolerant solar cells and light emitting diodes, our discoveries will accelerate more understanding and applications of perovskites.